\documentstyle[12pt]{article}
\newcommand{\be}{\begin{equation}}
\newcommand{\ee}{\end{equation}}
\newcommand{\ba}{\begin{eqnarray}}
\newcommand{\ea}{\end{eqnarray}}
\newcommand{\baa}{\begin{eqnarray*}}
\newcommand{\eaa}{\end{eqnarray*}}
\newcommand{\bb}{\begin{thebibliography}}
\newcommand{\eb}{\end{thebibliography}}
\newcommand{\ci}[1]{\cite{#1}}
\newcommand{\bi}[1]{\bibitem{#1}}
\newcommand{\lab}[1]{\label{#1}}
\newcommand{\re}[1]{(\ref{#1})}

\newcommand\fac[2]{\mbox{$\frac{#1}{#2}$}}

\newcommand\DGM{XX^{th} }

\begin{document}

\begin{titlepage}
\begin{flushright}
  UdeM-LPN-TH76  \\
   January 1992
\end{flushright}

\vspace*{30mm}
\begin{center}

{\LARGE Deformed Conformal and Supersymmetric \\

\medskip

Quantum Mechanics}

\vspace{10mm}

{\large Vyacheslav Spiridonov%
\footnote{On leave of absence from the Institute for Nuclear Research,
Moscow, Russia}$^,$%
\footnote{e-mail address: spiridonov@lps.umontreal.ca}}

\medskip

{\em Laboratoire de Physique Nucl\' eaire,
Universit\' e de Montr\' eal, \par
C.P. 6128, succ. A, Montr\' eal, Qu\' ebec, H3C 3J7, Canada}

\end{center}

\vspace*{12mm}
\begin{abstract}

Within the standard quantum mechanics a $q$-deformation of the
simplest $N=2$ supersymmetry algebra is suggested. Resulting physical
systems do not have conserved charges and degeneracies in the
spectra.
Instead, superpartner Hamiltonians are $q$-isospectral,
i.e. the spectrum of one can be obtained from another (with possible
exception of the lowest level) by the $q^2$-factor scaling.
A special class of the self-similar potentials is shown to obey
the dynamical conformal symmetry algebra $su_q(1,1)$.
These potentials exhibit exponential spectra and
corresponding raising and lowering operators
satisfy the $q$-deformed harmonic oscillator algebra
of Biedenharn and Macfarlane.

\medskip

\end{abstract}
\end{titlepage}

\newpage

\section{Introduction}

As a rule, group-theoretical methods alleviate description of the
complicated physical systems. In this respect Lie
algebras are of a special interest. Very elegant examples of their
applications were found in quantum mechanics within
the concept of the spectrum generating,
or, dynamical (super)symmetry algebras \ci{r1}.
Not long ago a wide attention was drawn to the deformations
of the Lie algebras known nowdays under the name of
quantum algebras, or, quantum groups \ci{r2}.
Physical models where a coupling
constant is related to the deformation parameter $q$ and where
Hamiltonian commutes with the generators of the quantum algebra
$su_q(2)$ were found on the linear lattice \ci{r3}. Thus,
some kind of equivalence of the perturbation of the
interaction between "particles" to the deformation of the symmetry
algebras governing the dynamics was demonstrated.

Biedenharn and Macfarlane introduced $q$-deformed harmonic oscillator
as a building block of the quantum algebras \ci{r4,r5}.
Many mathematical applications
of the $q$-oscillators appeared since that time \ci{r6,r7}
(an overview of the algebraic aspects of the $q$-analysis can be
found in Ref.\ci{r8}). Physical models of $q$-oscillators can be devided
into the three classes. The first class, inspired by Ref.\ci{r3},
is related to
the lattice systems \ci{r7,r9}. In the second class dynamical quantities
are defined on the "quantum planes" -- the spaces with non-commutative
coordinates \ci{r10}. Although Schr\" odinger equation in this approach
looks similar to the standard one, an explicit representation of it in
terms of the normal calculus results in the non-local finite-difference
equation. Parameter $q$ responsible for the non-commutativity of quantum
space coordinates serves as some non-local scale
on the continuous manifolds and, therefore, the basic physical principles
are drastically changed in this type of deformation.
We shall not pursue here the routes of these two groups of models.

The third -- dynamical symmetry realization class -- is purely
phenomenological: one deforms
already known spectra by choosing the Hamiltonian as some combination of
the quantum algebra generators \ci{r11}, or, as an anticommutator of the
formal $q$-oscillator creation and annihilation operators \ci{r4,r9}.
This application, in fact, does not have straightforward
physical meaning because of the non-uniqueness of the deformation
procedure. Even within the standard physical concepts exact knowledge
of the spectrum is not enough for the reconstruction of the potential.
For a given potential with infinite
number of bound states one can associate another potential with
infinitely many independent parameters and the same spectrum \ci{r12}.
Therefore the physics behind such deformations is not completely
fixed. One should precisely describe what kind of
interaction between the excitations leads to
peculiar change of the spectrum. Some analysis
of the inverse problems can be found in Ref.\ci{r13}, where
simulations of periodic potentials with the prescribed
band structure was performed, and in Ref.\ci{r14}, where a reconstruction of
artificial symmetric non-oscillating potential generating
prescribed discrete spectrum in the WKB-approximation was considered.
$q$-Analogs of the harmonic oscillators were also used for the
description of small violation of the
statistics of identical particles \ci{r15,r16}.

Recently the author proposed new approach to the problem of the
quantum algebra symmetries in physical models \ci{r17}, namely,
to take exactly solvable Schr\" odinger potentials and deform
their shape (e.g., by
changing the Taylor series expansion coefficients) in such a way that
the problem remains to be exactly solvable but the spectrum aquires
complicated functional character. This idea was stimulated by the
Shabat's one-dimensional reflectionless potential showing peculiar
self-similar behavior and describing an infinite
number soliton system \ci{r18}.
The latter was identified in Ref.\ci{r17} as a representative of a general
two parameter potential unifying via the $q$-deformation conformally
invariant harmonic and Coulomb, Rosen-Morse, and P\"oschl-Teller
potentials. The hidden quantum algebraic symmetry was claimed
to be responsible for the exponential character of the spectrum. In
comparison with the discussed above third group of models present
approach is the direct one -- physical interaction is fixed first
and the question on the quantum algebra behind prescribed
rule of $q$-deformation is secondary.

In this paper we extend further the results of Ref.\ci{r17} and propose
general deformation of the supersymmetric (SUSY) quantum mechanics
\ci{r19}. We define
$q$-SUSY algebra and provide explicit realization of that on the
Hilbert space of square integrable functions. The
degeneracies of standard SUSY models are lifted. 
The set of self-similar potentials naturally appears as a particular
example of $q$-SUSY system obeying dynamical symmetry algebra
$su_q(1,1)$. In particular, the raising and lowering operators
entering the definition of the supercharges are shown to generate
$q$-oscillator algebra of Biedenharn and Macfarlane.
The whole construction is based on the commutative analysis and has
many physical applications.

\section{SUSY quantum mechanics}

The simplest $N=2$ SUSY quantum mechanics is fixed by the following
algebraic relations between the Hamiltonian of a system $H$ and
supercharges $Q^{\dag} ,\, Q$ \ci{r19}
\be
\{Q,Q^{\dag}\}=H,\quad Q^2=(Q^{\dag})^2=0, \quad [H,Q]=[H,Q^{\dag}]=0.
\lab{e1}
\ee
All operators are supposed to be well defined on the relevant Hilbert
space. Then, independently on the explicit realizations the spectrum
is two-fold degenerate and the ground state energy is semipositive,
$E_{vac}\geq 0$.

Let us consider a particle moving in the one-dimensional space. Below,
the coordinate $x$ is tacitly assumed to cover the whole line,
$x\in {\it R}$, if it is not explicitly stated that it belongs to some cut.
Standard representation of the algebra \re{e1} contains one free
superpotential $W(x)$ \ci{r20}:
\be
Q=\left(\matrix{0&0\cr A&0\cr}\right), \quad
Q^{\dag}=\left(\matrix{0&A^{\dag}\cr 0&0\cr}\right), \quad
A=(p+iW(x))/\sqrt2,\quad [x,p]=i,
\lab{e2}
\ee
\be
H=\left(\matrix{H_+&0\cr 0&H_-\cr}\right)=
\left(\matrix{A^{\dag} A&0\cr 0&A A^{\dag}\cr}\right)
=\fac12(p^2+W^2(x)+W^\prime(x)\sigma_3),
\lab{e3}
\ee
$$W^\prime(x)\equiv {d\over dx} W(x), \qquad
\sigma_3=\left(\matrix{1&0\cr 0&-1\cr}\right).$$
It describes a particle with two-dimensional internal space the basis
vectors of which can be identified with the spin "up" and "down" states.

The subhamiltonians $H_\pm$ are isospectral as a result of the
intertwining relations
\be
H_+ A^{\dag}=A^{\dag} H_-,\qquad A H_+=H_- A.
\lab{e4}
\ee
The only possible difference concerns the lowest level. Note that the
choice $W(x)=-x$ corresponds to the harmonic oscillator problem and
then $A^{\dag},\, A$ coincide with the bosonic creation and
annihilation operators $a^{\dag},\,a$ which satisfy the algebra
\be
[a,a^{\dag}]=1,\qquad [N,a^{\dag}]=a^{\dag},\qquad [N,a]=-a,
\lab{e5}
\ee
where $N$ is the number operator, $N=a^{\dag} a$. This, and another
particular choice, $W(x)=\lambda/x$, correspond to the conformally
invariant dynamics \ci{r21}.

\section{$q$-Deformed SUSY quantum mechanics}

Now we shall introduce the tools needed for the quantum algebraic
deformation of the above construction. Let $T_q$ be smooth $q$-scaling
operator defined on the continuous functions
\be
T_q f(x)=f(qx),
\lab{e6}
\ee
where $q$ is a real non-negative parameter. Evident properties of this
operator are listed below
$$ T_q f(x)g(x)=[T_q f(x)][T_q g(x)],\qquad
T_q {d\over dx}=q^{-1}{d\over dx} T_q, $$
\be
T_q T_p=T_{qp},\qquad T^{-1}_q=T_{q^{-1}},\qquad T_1=1.
\lab{e7}
\ee
On the Hilbert space of square integrable functions ${\cal L}_2$ one has
\be
\int_{-\infty}^{\infty} \phi^*(x)\psi(qx)dx=
q^{-1}\int_{-\infty}^{\infty} \phi^*(q^{-1}x)\psi(x)dx,
\lab{e8}
\ee
where from the hermitian conjugate of $T_q$ can be found
\be
T_q^{\dag}=q^{-1} T_q^{-1},\qquad \quad  (T_q^{\dag})^{\dag}=T_q.
\lab{e9}
\ee
As a result, $\sqrt{\, q}\, T_q$ is a unitary operator.
Because we take wave functions to be infinitely differentiable,
an explicit realization of $T_q$ is provided by the operator
\be
T_q=e^{\ln q\, x\,d/dx}=q^{x\,d/dx}.
\lab{e10}
\ee
Expanding \re{e10} into the formal series and using integration by parts
one can prove relations \re{e9} on the finite coordinate cut as well because
wave functions vanish on the boundaries.

Let us define the $q$-deformed factorization operators
\be
A^{\dag}={1\over \sqrt2}\, (p-iW(x))\,T_q, \qquad
A={q^{-1}\over \sqrt2}\, T_q^{-1} (p+iW(x)),
\lab{e11}
\ee
where $W(x)$ is arbitrary function and for the convinience we use
the same notations as in the undeformed case \re{e3}. $A$ and $A^{\dag}$
are hermitian conjugates of each other on the ${\cal L}_2$. Now one has
\ba
A^{\dag} A&=&\fac12 q^{-1} (p^2+W^2(x)+W^\prime(x))\equiv q^{-1} H_+,
\lab{e12}   \\
A\, A^{\dag}&=&\fac12q^{-1} T_q^{-1}(p^2+W^2(x)- W^\prime(x))T_q
\nonumber \\
&=&\fac12q\,(p^2+q^{-2}W^2(q^{-1}x) - q^{-1}W^\prime (q^{-1}x))
\equiv q H_-.
\lab{e13}
\ea
We define $q$-deformed SUSY Hamiltonian and supercharges to be
\be
H=\left(\matrix{H_+&0\cr 0&H_-\cr}\right)
=\left(\matrix{qA^{\dag} A&0\cr 0&q^{-1}A A^{\dag}\cr}\right),\qquad
Q=\left(\matrix{0&0\cr A&0\cr}\right),\quad
Q^{\dag}=\left(\matrix{0&A^{\dag}\cr 0&0\cr}\right).
\lab{e14}
\ee
These operators satisfy the following $q$-deformed version
of the $N=2$ SUSY algebra
\be
\{Q^{\dag},Q\}_q= H, \quad \{Q,Q\}_q=\{Q^{\dag},Q^{\dag}\}_q=0,\quad
[H,Q]_q=[Q^{\dag}, H]_q=0,
\lab{e15}
\ee
where we introduced $q$-brackets
\be
[X,Y]_q\equiv qXY-q^{-1}YX,\qquad [Y,X]_q=-[X,Y]_{q^{-1}},
\lab{e16}
\ee
\be
\{X,Y\}_q\equiv qXY+q^{-1}YX,\qquad \{Y,X\}_q=\{X,Y\}_{q^{-1}}.
\lab{e17}
\ee
Note that the supercharges are not conserved because they do not commute
with the Hamiltonian (in this respect our algebra principally
differs from the formal construction of Ref.\ci{r22}). An interesting
property of the algebra \re{e15} is that it shares with \re{e1}
the semipositiveness of the ground state energy which follows from the
observation that $Q^{\dag},\, Q$ and the operator $q^{-\sigma_3} H$
satisfy ordinary SUSY algebra \re{e1}. Evidently,
in the limit $q\to 1$ one recovers conventional SUSY quantum mechanics.

For the subhamiltonians $H_\pm$ the intertwining relations look as
follows
\be
H_+ A^{\dag}=q^2 A^{\dag} H_-,\qquad A H_+=q^2 H_- A.
\lab{e18}
\ee
Hence, $H_\pm$ are not isospectral but rather $q$-isospectral,
i.e. the spectrum of $H_+$ can be obtained from the spectrum of
$H_-$ just by the $q^2$-factor scaling:
\be
H_+\, \psi^{(+)}=E^{(+)}\psi^{(+)}, \qquad
H_-\, \psi^{(-)}=E^{(-)}\psi^{(-)},
\nonumber
\ee
\be
E^{(+)}=q^2\, E^{(-)}, \qquad
\psi^{(+)}\propto A^{\dag} \psi^{(-)}, \quad
\psi^{(-)}\propto A\, \psi^{(+)}.
\lab{e18a}
\ee
Possible exception concerns only the lowest level in the same spirit
as it was in the undeformed SUSY quantum mechanics. If $A^{\dag}, A$
do not have zero modes then there is one-to-one correspondence between
the spectra. We name this situation as a spontaneously broken $q$-SUSY
because for it $E_{vac}>0$. If $A$ (or, $A^{\dag}$) has zero mode
then $q$-SUSY is exact, $E_{vac}=0$, and $H_-$ (or, $H_+$) has one level
less than its superpartner $H_+$ (or, $H_-$).

As a simplest physical example let us consider the case $W(x)=-x$. The
Hamiltonian takes the form
\be
4H=2p^2+(1+q^{-4})x^2 +q^{-2}-1+((1-q^{-4})x^2-1-q^{-2})\sigma_3,
\lab{e19}
\ee
and describes a spin-1/2 particle in the harmonic well and
related magnetic field along the third axis.
The physical meaning of the deformation parameter $q$ is analogous
to that in the XXZ-model \ci{r3} -- it is a specific interaction constant
in the standard physical sense. This model has exact $q$-SUSY
and if $q^2$ is equal to the rational number
the spectrum exhibits accidental degeneracies.

\section{General deformation of the SUSY quantum mechanics}

Described above $q$-deformation of the SUSY quantum mechanics is by
no means unique. If one choses in the formulae \re{e11} $T_q$ to be not
$q$-scaling operator but, instead, the shift operator
\be
T_q f(x)=f(x+q), \qquad T_q=e^{q\, d/dx},
\lab{e20}
\ee
then SUSY algebra will not be deformed at all. The superpartner
Hamiltonians will be isospectral and the presence of the
$T_q$-operator results in the very simple deformation of the
standard superpartner potential $U_-(x)\to U_-(x-q)$ (kinetic
term is invariant). Evidently
such deformation does not change the spectrum of $U_-(x)$ and
that is why SUSY algebra remains intact but physically this
creates new SUSY quantum mechanical models. The crucial
point in generating of them was the use of the essentially
infinite order differential operators as the intertwining operators.

The most general choice of $T_q$ would be the shift operator
in the arbitrarily chosen change of the variable function $z=z(x)$
\be
T_q f(z(x))= f(z(x)+q), \qquad
T_q=e^{q\,d/dz(x)},\quad {d\over dz}={1\over z^\prime(x)}\,
{d\over dx}.
\lab{e21}
\ee
The choices $z=\ln x$ and $z=x$ were already discussed above.
In general, operator $T_q$ will not preserve the form of
the kinetical term in the $H_-$-Hamiltonian. Physically such change
would correspond to the transition from the motion of the particle
on the flat space to the curved space dynamics.
Application of the described construction to the spherically symmetric
potentials is straightforward. The general higher dimensional SUSY
models are more complicated but can also be "deformed".
It is evident that all quantum mechanical problems discussed within
the SUSY approach can be considered in the suggested fasion.
We leave detailed discussion of the approach and physical
applications for the future publications.

\section{$q$-Deformed conformal quantum mechanics}

Explicit form of the $su(1,1)$ dynamical symmetry generators can be
read off from the harmonic oscillator \re{e5} realization
\be
K_+=\fac12 (a^{\dag})^2,\qquad K_-=\fac12 a^2,\qquad
K_0=\fac12 (N+\fac12),
\lab{e22}
\ee
\be
[K_0, K_\pm]=\pm \,K_\pm,\qquad [K_+,K_-]=-2K_0.
\lab{e23}
\ee
Let us show that the potentials considered in Refs.\ci{r18,r17}
obey the quantum conformal symmetry algebra $su_q(1,1)$.

First, we shall derive those potentials within another physical
situation with the help of $q$-SUSY. Let us consider the Hamiltonian
of a spin-1/2 particle in the external potential $\fac12 U(x)$ and
the magnetic field $\fac12 B(x)$ along the third axis
\be
H=\fac12 (p^2 + U(x) + B(x) \sigma_3)
\lab{e24}
\ee
and impose two conditions: we take magnetic field to be homogeneous
\be
B=-\beta^2 q^{-2} = const
\lab{e25}
\ee
and require the presence of $q$-SUSY \re{e15}. Equating \re{e24}
and \re{e14} we arrive at the potential
\be
U(x)=W^2(x)+W^{\prime}(x) + \beta^2 q^{-2},
\lab{e26}
\ee
where $W(x)$ satisfies the following mixed finite-difference and
differential equation
\be
W^\prime(x)-W^2(x)+qW^\prime (qx)+q^2 W^2(qx)+2\beta^2 =0.
\lab{e27}
\ee
This is the condition of a self-similarity \ci{r18,r17} which bootstraps
the potential in different points (in Ref.\ci{r17}
$\beta^2= \gamma^2 (1+q^2)/2$ parametrization was used).
Smooth solution of \re{e27} for symmetric potentials
$U(-x)=U(x)$ is given by the following power series
\be
W(x)=\sum_{i=1}^{\infty} c_i\, x^{2i-1}, \qquad
c_i={1-q^{2i}\over 1+q^{2i}}{1\over 2i-1}\sum_{m=1}^{i-1}c_{i-m}c_m, \quad
c_1=-\, {2\beta^2\over 1+q^2}.
\lab{e27a}
\ee
In different limits of the parameters several well known
exactly solvable problems arise: 1. Rosen-Morse -- at $q\to 0;\;$
2. P\"oschl-Teller -- at $\beta\propto q\to \infty ;\;$ 3. Harmonic well --
at $q\to 1;\;$ 4. Coulomb potential -- at $q\to 0$ and $\beta = 0$.
Note that at $q>1$ the range of the coordinate should be
restricted to the finite cut. Soliton solution of Shabat \ci{r17}
corresponds to the range $0<q<1$ at fixed $\beta$.
Within the taken physical situation the potential \re{e26}
describes $q$-deformation of the Landau-level problem.

We already know that the spectra of $H_\pm$ subhamiltonians are
related via the $q^2$-scaling
\be
E^{(+)}_{n+1}=q^2 E^{(-)}_n,
\lab{e27b}
\ee
where the number $n$ numerates levels from below for both spectra.
Because $q$-SUSY is exact in this model the lowest level of $H_-$
corresponds to the first excited state of $H_+$. But at the restriction
\re{e25} the spectra differ only by a constant,
\be
E^{(+)}_n=E^{(-)}_n -\beta^2 q^{-2},
\lab{e27c}
\ee
Conditions \re{e27b} and \re{e27c} give us the spectrum
\be
E_{n,m}=\beta^2 \, {q^{-2m}-q^{2n}\over 1-q^2},
\qquad   m=0,1;\; n=0,1,\dots,\infty .
\lab{e28}
\ee
At $q<1$ there are two finite accumulation points, i.e. \re{e28} somehow
approximates the two-band spectrum. At $q > 1$ energy eigenvalues
exponentially grow to the infinity.

Not more difficult is the derivation of the dynamical symmetry algebra.
To find that we rewrite relations \re{e12}, \re{e13} for the
superpotential \re{e27a}
\be
A^{\dag}A=q^{-1} H+{\beta^2 q^{-1}\over 1-q^2}, \qquad
A\, A^{\dag}=q\, H+{\beta^2 q^{-1}\over 1-q^2},
\lab{e29}
\ee
where H is the Hamiltonian with purely exponential spectrum
\be
H=\fac12 (p^2+W^2(x)+W^{\prime}(x))- {\beta^2 \over 1-q^2},\qquad
E_n=-{\beta^2 \over 1-q^2}\, q^{2n}.
\lab{e29a}
\ee
Evidently,
\be
AA^{\dag}-q^2 A^{\dag}A=\beta^2 q^{-1}.
\lab{e30}
\ee
Normalization of the r.h.s. of \re{e30} to unity results in the
algebra used in Ref.\ci{r16} for the description of small violation
of the Bose statistics.

The shifted Hamiltonian \re{e29a} and $A^{\dag},\, A$ operators $q$-commute
$$[A^{\dag},H]_q=[H,A]_q=0,$$
or,
\be
H\, A^{\dag}=q^2A^{\dag} H,\qquad A\, H=q^2H\, A.
\lab{e30a}
\ee
These are typical braid-type commutation relations.
Energy eigenfunctions $| n\rangle $ can be uniquely determined
from the ladder operators action
\be
A^{\dag}|n\rangle =\beta q^{-1/2}\sqrt{\,{1-q^{2(n+1)}\over 1-q^2}}\,
|n+1\rangle ,\qquad
A\, |n\rangle =\beta q^{-1/2}\sqrt{\, {1-q^{2n}\over 1-q^2}}\,
|n-1\rangle .
\lab{e30b}
\ee

It is convinient to introduce the formal number operator
\be
N={\ln [(q^2-1) H/\beta^2]\over \ln q^2},\qquad
N\, |n\rangle =n |n\rangle,
\lab{e31}
\ee
which is well defined only on the eigenstates of the Hamiltonian.
Now one can check that the operators
\be
a_q={q\over \beta}\,A\, q^{-N/2},\qquad
a^{\dag}_q={q\over \beta}\, q^{-N/2} A^{\dag}
\lab{e32}
\ee
satisfy the $q$-deformed harmonic oscillator algebra
of Biedenharn and Macfarlane \ci{r4,r5}
\be
a_q a^{\dag}_q - q a^{\dag}_q a_q=q^{-N},\quad
[N,a^{\dag}_q]=a^{\dag}_q,\quad [N,a_q]=-a_q.
\lab{e33}
\ee
Substituting $a^{\dag}_q$ and $a_q$ into the definitions \re{e22} and
renormalizing $2K_{\pm}/(q+q^{-1})\to K_{\pm}$ we get commutation
relations of the quantum algebra $su_q(1,1)$
\be
[K_0,K_\pm]=\pm\, K_\pm,\qquad [K_+, K_-]=-\,
{\mu^{2K_0}-\mu^{-2K_0}\over \mu - \mu^{-1}},\quad  \mu=q^2.
\lab{e34}
\ee
Therefore the dynamical symmetry algebra of the model is $su_q(1,1)$.

Let us compare our deformed (super)conformal quantum mechanics with the
construction of Ref.\ci{r23}. Kalnins, Levine, and Miller
called as the conformal symmetry generator any differential operator
$L(t)$ which maps solutions of the
time-dependent Schr\"odinger equation to the solutions, i.e. which
satisfies the relation
\be
i\, {\partial \over \partial t}\,  L-[H,L]=
R\, (i\, {\partial \over \partial t}-H),
\lab{e35}
\ee
where $R$ is some operator. On the shell of Schr\" odinger equation
solutions $L(t)$ is conserved and all higher
powers of the space derivative, entering the definition of $L(t)$,
can be replaced by the powers of $\partial /\partial t$ and
linear in $\partial /\partial x$ term. But any analytical
function of $\partial /\partial t$ is replaced by the
function of energy when applied to the stationary states.
This trick allows
to simulate any infinite order differential operator by the one
linear in space derivative and to prove that a solution with energy
$E$ can always be mapped to the not-necessarily normalizable solution with
the energy $E+f(E)$ where $f(E)$ is arbitrary analytical function.
"On-shell" raising and lowering operators always can be
found if one knows the basis solutions of the
Schr\"odinger equation but sometimes it is easier
to find symmetry generators and use them in search of the spectrum.

In our construction we have "off-shell" symmetry generators, which
satisfy quantum algebraic relations in the operator sense.
In this respect our results are complimentary to those of the
Ref.\ci{r23}. Indeed, time-dependent factorization operators
\be
{\cal A}^{\dag}(t)=e^{i(q^{-2}-1)tH} A^{\dag}, \qquad
{\cal A}(t)=e^{i(q^2-1)tH} A
\lab{e36}
\ee
satisfy \re{e35} with $R\equiv 0$ and, so, are real conserved
quantities. On the stationary states the exponential prefactors
in \re{e36} coincide with the time shift operators. One can reduce
operators ${\cal A}^{\dag}(t),\, {\cal A}(t)$ to the linear in
$\partial /\partial x$ "on-shell"-form and then they, probably,
shall coincide with the ladder operators of Ref.\ci{r23} corresponding
to the Hamiltonian \re{e29a}.

\section{Conclusions}

To conclude, in this paper we have suggested a deformation
of the  SUSY quantum mechanics. The main feature of the construction
is that the superpartner Hamiltonians satisfy non-trivial braid-type
intertwining relations which remove degeneracies of the original SUSY
spectra. Deformed SUSY algebra preserves semipositiveness of the
vacuum energy. Peculiar set of $q$-SUSY potentials arising within
the Landau-level-like problem obey $q$-deformed dynamical
conformal symmetry algebra $su_q(1,1)$. Corresponding raising and
lowering operators satisfy $q$-deformed oscillator algebra
of Biedenharn and Macfarlane. A more general type of potential
deformations applicable in any dimensional space is outlined.

It is clear that $q$-scaling is a particular example of the possible
transformations of the spectra.
In general one should be able to analytically describe the map
of a given potential with spectrum $E_n$ to the particular potential
with the spectrum $f(E_n)$ for any analytical function $f(E)$.
A problem of arbitrary non-linear deformation of the Lie algebras
was treated in Ref.\ci{r24} using the symbols of operators
without well defined
coordinate representaion on the ordinary Hilbert space ${\cal L}_2$.
Certainly, the method of Ref.\ci{r23} should be
helpful in the analysis of this interesting problem and the
model with exponential spectrum given in Sect. 5 shows that sometimes
one can even find well defined "off-shell"  spectrum generating algebra.

The Hopf algebra structure of the quantum groups was not mentioned
because its physical meaning within the standard quantum mechanical
context is unknown to the author. Perhaps the many identical body
problems shall elucidate this point. Another speculative conjecture
is that the results of this and \ci{r17,r18} papers may
be useful in seeking for the $q$-deformations of the non-linear
integrable evolution equations, like KdV, {\it sin}-Gordon, etc.
We end by the remark that presented type of $q$-deformation can also
be developed for the parasupersymmetric quantum mechanics \ci{r25}
where higher (odd and even) dimensional internal spaces are involved.

\bigskip\bigskip
\noindent{\Large{\bf Acknowledgments}}
\bigskip

The author is indebted to W.Miller and A.Shabat for the useful
discussions on the initial stage of this work, to J.LeTourneux and
L.Vinet for the encuragement to pursue the subject.
This research is supported by the NSERC of Canada.

\newpage

\bb{33}

\bi{r1} Dynamical Groups and Spectrum Generating Algebras,
        Eds. A.Bohm, Y.Ne'eman, and A.Barut (World Scientific, 1988).

\bi{r2} V.G.Drinfeld, Quantum Groups, {\it in:} Proc. of the Intern.
      Congress of Mathematicians (Berkeley, 1986) vol.1, p.798;   \\
        M.Jimbo, Lett.Math.Phys. {\bf 10} (1985) 63; {\bf 11}
        (1986) 247; \\ N.Yu.Reshetikhin, L.A.Takhtajan, and
        L.D.Faddeev, Algebra i Analiz, {\bf 1} (1989) 178.

\bi{r3} N.Pasquier and H.Saleur, Nucl.Phys. {\bf B330} (1990) 523.

\bi{r4} L.C.Biedenharn, J.Phys. {\bf A22} (1989) L873.

\bi{r5} A.J.Macfarlane, J.Phys. {\bf A22} (1989) 4581.

\bi{r6} C.-P. Sen and H.-C.Fu, J.Phys. {\bf A22} (1989) L983; \\
        T.Hayashi, Comm.Math.Phys. {\bf 127} (1990) 129;  \\
        M.Chaichian and P.Kulish, Phys.Lett. {\bf B234} (1990) 72; \\
        P.P.Kulish and E.V.Damaskinsky, J.Phys. {\bf A23} (1990) L415; \\
        R.Floreanini, V.P.Spiridonov, and L.Vinet, Comm.Math.Phys.
        {\bf 137} (1991) 149;  \\
        D.B.Fairlie and C.Zachos, Quantized Planes and Multiparameter
        Deformations of Heisenberg and $GL(N)$ Algebras, preprint
        ANL-HEP-CP-91-28, 1991; \\
      R.Floreanini, D.Leites, and L.Vinet, On the Defining Relations
      of the Quantum Superalgebras, preprint UdeM-LPN-TH53, 1991.

\bi{r7} N.M.Atakishiev and S.K.Suslov, Sov.J.Theor.Math.Phys.
        {\bf 85} (1990) 1055.

\bi{r8} R.Floreanini and L.Vinet, Representations of Quantum Algebras
        and $q$-Special Functions, {\it in:} Proc. of the ${\it II^{nd}}$
        Intern. Wigner Symposium (Springer-Verlag, 1991); preprint
        UdeM-LPN-TH69, 1991.

\bi{r9} E.G.Floratos and T.N.Tomaras, Phys.Lett. {\bf B251} (1990) 163.

\bi{r10}J.Wess and B.Zumino, Nucl.Phys. (Proc.Suppl.) {\bf B18}
        (1990) 302; \\ B.Zumino, Mod.Phys.Lett. {\bf A6} (1991) 1225; \\
        U.Carow-Watamura, M.Schlieker, and S.Watamura, Z.Phys.
        {\bf C49} (1991) 439;  \\
        J.A.Minanhan, Mod.Phys.Lett. {\bf A5} (1990) 2635;  \\
        L.Baulieu and E.G.Floratos, Phys.Lett. {\bf B258} (1991) 171.

\bi{r11} P.P.Raychev, R.P.Roussev, and Y.F.Smirnov, J.Phys. {\bf G16}
         (1990) L137; \\
         M.Chaichian, D.Ellinas, and P.Kulish, Phys.Rev.Lett. {\bf 65}
         (1990) 980; \\
         R.M.Mir-Kasimov, The Relativistic Oscillator as the Realization
         of the Quantum Group of Dynamical Symmetry, {\it in:} Proc.
         of the Intern. Seminar "Quarks'90", 14-19 May 1990, Telavi,
     USSR. Eds. V.A.Matveev et al (World Scientific, Singapore) p.133; \\
         E.G.Floratos, J.Phys. {\bf A24} (1991) 4739.

\bi{r12} M.M.Nieto, Phys.Lett. {\bf B145} (1984) 208; \\
         M.Luban and D.L.Pursey, Phys.Rev. {\bf D33} (1986) 431.

\bi{r13} J.L.Rosner, Ann.Phys.(N.Y.) {\bf 200} (1990) 101.

\bi{r14} D.Bonatsos, C.Daskaloyannis, and K.Kokkotas,
         J.Phys. {\bf A24} (1991) L795.

\bi{r15} A.Yu.Ignatiev and V.A.Kuzmin, Yad.Fiz. {\bf 46} (1987) 786.

\bi{r16} O.W.Greenberg, Phys.Rev.Lett. {\bf 64} (1990) 705; \\
        R.Mohapatra, Phys.Lett. {\bf B242} (1990) 407;  \\
        V.P.Spiridonov, Dynamical Parasupersymmetry in Quantum
        Systems, {\it in:} Proc. of the Intern. Seminar "Quarks'90",
        14-19 May 1990, Telavi, USSR. Eds. V.A.Matveev et al (World
        Scientific, Singapore) p.232

\bi{r17} V.Spiridonov, Exactly Solvable $q$-Deformed Potentials with
        Exponentially Small or Large Bound Energies, preprint
        UdeM-LPN-TH75, 1991.

\bi{r18} A.Shabat, The Infinite-Dimensional Dressing Dynamical System;
         Inverse Problems, to be published.

\bi{r19} L.E.Gendenstein and I.V.Krive, Sov.Phys.Usp. {\bf 28} (1985) 645.

\bi{r20} E.Witten, Nucl.Phys. {\bf B188} (1981) 513.

\bi{r21} V.DeAlfaro, S.Fubini, and G.Furlan, Nuovo Cim. {\bf A34} (1976) 569.

\bi{r22} M.Chaichian, P.Kulish, and J.Lukierski, Phys.Lett.
        {\bf B262} (1991) 43.

\bi{r23} E.G.Kalnins, R.D.Levine, and W.Miller, Jr., Conformal Symmetries
        and Generalized Recurrences for Heat and Schr\"odinger
        Equations in One Spatial Dimension, {\it in:}
        Mechanics, Analysis and Geometry: 200 Years after Lagrange,
        Ed. M.Francaviglia (Elsevier Science Publishers B.V., 1991) p. 237

\bi{r24} A.P.Polychronakos, Mod.Phys.Lett. {\bf A5} (1990) 2325;  \\
         M.Ro{\v c}ek, Phys.Lett. {\bf B255} (1991) 554;  \\
         K.Odaka, T.Kishi, and S.Kamefuchi, J.Phys. {\bf A24} (1991) L591; \\
         C.Daskaloyannis, J.Phys. {\bf A24} (1991) L789;  \\
         C.Daskaloyannis and K.Ypsilantis, A Deformed Oscillator with
         Coulomb Energy Spectrum, preprint THES-TP-91/09, 1991.

\bi{r25} V.A.Rubakov and V.P.Spiridonov, Mod. Phys. Lett. {\bf A3}
         (1988) 1337; \\ V.Spiridonov, Parasupersymmetry in Quantum
         Systems, {\it in:} Proc. of the $\DGM$ Intern. Conf. on the Diff.
         Geometry Methods in Theor. Physics (World Scientific, 1991);
         preprint UdeM-LPN-TH58, 1991.

\eb
\end{document}